\documentclass[preprint,12pt]{elsarticle}

\usepackage{amssymb,amsmath,amsthm,url}

\begin{document}
\newtheorem{theorem}{Theorem}
\newtheorem{remark}{Remark}
\begin{frontmatter}
\title{Cryptanalysis and improvement of two  certificateless three-party  authenticated key agreement protocols} 
\author{Haiyan Sun \corref{a}}\ead{wenzhong2520@gmail.com}
\author{Qiaoyan Wen}
\author{Hua  Zhang}
\author{Zhengping Jin}
\author{Wenmin Li}
\cortext[a]{Corresponding author.}
\address{State Key Laboratory of Networking and Switching
Technology,\\
Beijing University of Posts and Telecommunications, Beijing 100876,
China}
\begin{abstract}
 Recently, two certificateless three-party  authenticated key agreement protocols were  proposed,  and  both protocols were claimed they can meet the desirable security properties including forward security, key compromise impersonation resistance and so on. Through cryptanalysis, we show that one  neither meets forward security and key compromise impersonation resistance nor resists an attack by an adversary who knows all users' secret values, and the other cannot resist key compromise impersonation attack. Finally, we propose improved protocols to make up two original protocols' security weaknesses, respectively. Further security analysis shows that our  improved protocols can remove such security weaknesses.
\end{abstract}

\end{frontmatter}

\section{Introduction}

Authenticated key agreement (AKA) is one of the fundamental cryptographic primitives. It allows two or more users to generate a shared session
secret key over an open network with each other, and all the users are assured that only their intended peers can know the shared session
secret key. AKA protocols can be realized in the traditional public-key infrastructure (PKI) setting, identity-based cryptography
setting \cite{Shamir}, or certificateless cryptography setting \cite{ALP}.   Certificateless authenticated key agreement (CLAKA)  protocols
would be more appealing due to its advantages in eliminating the heavy certificate management burden in PKI-based AKA protocols and  key escrow
problem in identity-based AKA protocols.  By far, many researchers have been investigating secure and efficient certificateless two-party authenticated key agreement protocols (e.g., \cite{WCD,WZ,SJ,LBN,ZZWD,HCC,HCCZH,XQC,YT,HPC}). A research direction in AKA protocol aims to generalize two-party AKA setting to multi-party AKA setting, among which the three-party AKA protocols receive much interest. In 2009, Gao
et al. \cite{GZG} proposed the first three-party CLAKA protocol. Since then, several three-party CLAKA protocols (e.g., \cite{HXG}, the XCQ-11 protocol \cite{XCQ}, the XCL-12 protocol \cite{XCL}) have been proposed.

In this paper,  we analyze two three-party CLAKA protocols \cite{XCQ,XCL} and propose two improved protocols. Firstly, we point out that the XCQ-11 protocol \cite{XCQ} is subjected to three attacks including forward security attack, key compromise impersonation attack, and an attack by adversaries who know all users' secret values,   and  then propose a simple improvement to remove these flaws. Secondly, we find that the XCL-12 protocol \cite{XCL} cannot resist key compromise impersonation attack and  propose an efficient protocol which can resist this attack.

The remaining part of this paper is organized as follows. Some preliminaries are introduced in Section 2. A review and three attacks and an improved  protocol of  the XCQ-11 protocol are given in Section 3. A review and two attacks and an improved protocol of  the XCL-12 protocol are given in Section 4.   Finally, some conclusions are drawn in Section 5.
\section{Preliminaries}
We now briefly review some basic concepts used in this paper, including bilinear pairings and some security properties.
\subsection{Bilinear pairing}
Let $\mathbb{G}_1$ be an additive group generated by $P$ with prime order $q$ and  $\mathbb{G}_2$ be a multiplicative group of the same
order. A map $\hat{e}: \mathbb{G}_1\times \mathbb{G}_1\rightarrow \mathbb{G}_2$ is said to be a bilinear pairing if the
following three conditions hold true:
\leftmargini=4mm
\leftmarginii=4mm
\begin{enumerate}
 \item \verb  Bilinearity: for all $a,b\in\mathbb{Z}_q^*$, $\hat{e}(aP, bP)=\hat{e}(P, P)^{ab}$.

 \item \verb  Non-degeneracy: $\hat{e}(P, P)\neq 1_{\mathbb{G}_2}$.

 \item \verb  Computability: $\hat{e}$ is efficiently computable.
\end{enumerate}
\subsection{Security properties}

It is desirable for three-party authenticated key agreement protocols to possess the following security properties. Let $A$, $B$ and $C$ be three participants that execute the protocol correctly.
\leftmargini=3mm
\leftmarginii=2mm
\begin{itemize}

\item   {\bf Known-key security}: The session key is not compromised in the face of adversaries who have learned some other session keys.

\item   {\bf Key compromise impersonation  (KCI) resistance}: If an adversary reveals $A$'s long-term private key, the adversary cannot impersonate any other participant to $A$ without the participant's private key.

\item  {\bf Forward secrecy (FS)}: Compromising of long-term private keys of one or more of the  participants should not affect the secrecy of previously established session keys. A protocol has forward secrecy if the secrecy of previously established session keys is not affected when some but not all of the participants' long-term private keys are corrupted. A protocol has perfect forward secrecy if the secrecy of previously established session keys is not affected when all participants' long-term private keys are compromised.

\item  {\bf Unknown key share (UKS) resistance}:
    If one participant $A$ thinks that he/she is sharing a key
with the other participants (e.g., $B$ and $C$), then it should not happen that $A$ is actually
sharing that key with the adversary, which is not $B$ or $C$.
\end{itemize}
\section{The XCQ-11 protocol and its analysis and improvement}
In this section, we first review the XCQ-11 protocol \cite{XCQ}, then give three attacks on the XCQ-11 protocol, and finally  propose a simple countermeasure to resist these attacks.
\subsection{Review of  the XCQ-11 protocol}
The XCQ-11 protocol \cite{XCQ} requires a KGC and consists of four phases: system setup, partial key extraction, user key generation and key agreement phases.
\leftmargini=3mm
\leftmarginii=5.5mm
\begin{itemize}
\setlength{\parskip}{-1pt}
\item  {\bf Setup}: Given a security parameter $k\in \mathbb{Z}$, the algorithm works as follows.
\begin{enumerate}
\setlength{\parskip}{-3pt}
\item[(1)] It runs the parameter generator on input $k$ to generate a prime $q$, two groups $\mathbb{G}_1, \mathbb{G}_2$ of prime order $q$, a generator $P$ of $\mathbb{G}_1$, and an admissible pairing $\hat{e}: \mathbb{G}_1 \times\mathbb{G}_1 \rightarrow \mathbb{G}_2$.

\item[(2)]  It chooses a master-key $x\in Z_q^*$ and computes $P_0 = xP$.

\item[(3)]  It chooses three cryptographic secure hash functions $H_1: \{0,1\}^*  \rightarrow \mathbb{Z}_q^*$, $H_2: \mathbb{G}_1 \rightarrow \mathbb{Z}_q^*$  and $H_3:{\{0,1\}^*}^3 \times  \mathbb{G}_1^{9}\times \mathbb{G}_2  \rightarrow \{0,1\}^k$. Finally the KGC's master-key $x$ is kept secret and the system parameters $\{q,\mathbb{G}_1,\mathbb{G}_2, \hat{e}, P, P_0, H_1, H_2,H_3\}$ are published.
\end{enumerate}

\item {\bf PartialKeyGen}: Given a user's identity $ID_U \in \{0,1\}^*$, KGC first chooses at random $q_{U}=H_1(ID_U)$. It then sets this user's partial private key $s_U=\frac{1}{x+q_U} P$ and transmits it to user $ID_U$ secretly.

It is easy to see $s_U$ is actually a signature on $ID_U$ for the key pair $(P_0,x)$, and user $ID_U$ can check its correctness by checking whether $\hat{e}(s_U, P_0 + q_UP) = \hat{e}(P,P)$. For convenience, here we define $Q_U= P_0+ q_UP$.

\item {\bf UserKeyGen}: User  $ID_U$  picks randomly $x_U\in Z_q^*$ as his/her user secret key $usk_U$, and computes  his/her public key as $upk_U = x_UQ_U$. After that, the user $ID_U$ computes the full private key $S_U= \frac{1}{x_U+H_2(upk_U)}s_U$.

\item {\bf Key Agreement}: Assume that an entity $A$ with identity $ID_A$ has full private key $S_A$ and public key $upk_A$, an entity $B$ with identity $ID_B$ has private key $S_B$ and public key $upk_B$, and an entity $C$ with identity $ID_C$ has private key $S_C$ and public key $upk_C$. The message flows and computations of a protocol run are described below.
\begin{enumerate}
\setlength{\parskip}{0pt}
\item[(1)] $A, B, C$: choose $a, b, c \in Z_q^*$.
\item[(2)] $A \rightarrow B: T_{AB}= a(upk_B + H_2(upk_B)Q_B),$

\noindent $A \rightarrow C: T_{AC}= a(upk_C + H_2(upk_C)Q_C),$

\noindent $B \rightarrow A: T_{BA}= b(upk_A + H_2(upk_A)Q_A),$

\noindent $B \rightarrow C: T_{BC}= b(upk_C + H_2(upk_C)Q_C),$

\noindent $C \rightarrow A: T_{CA}= c(upk_A + H_2(upk_A)Q_A),$

\noindent $C \rightarrow B: T_{CB}= c(upk_B + H_2(upk_B)Q_B).$

\item[(3)]
$A: k_A =\hat{e}(P, P)^a \hat{e}(T_{BA}, S_A)\hat{e}(T_{CA}, S_A)=  \hat{e}(P, P)^{a+b+c},$

\noindent $B: k_B =\hat{e}(P, P)^b \hat{e}(T_{AB}, S_B)\hat{e}(T_{CB}, S_B)=  \hat{e}(P, P)^{a+b+c},$

\noindent $C: k_C =\hat{e}(P, P)^c \hat{e}(T_{AC}, S_C)\hat{e}(T_{BC}, S_C)=  \hat{e}(P, P)^{a+b+c}.$

\end{enumerate}
\end{itemize}

After the protocol has finished, all three entities share the session key, which is computed as

$K=H_3(ID_A||ID_B||ID_C||upk_A||upk_B||upk_C||T_{AB}||T_{AC}||T_{BA}||T_{BC}||T_{CA}||T_{CB}||$

$ \qquad \ \ \hat{e}(P, P)^{a+b+c})$.
\subsection{Failure to provide  forward secrecy}
Suppose that $A$'s long-term private key $S_A$ and $B$'s long-term private key $S_B$  have been compromised. In the following, we  show that an adversary $\mathcal{A}$ with the knowledge $S_A$ and $S_B$ can obtain previously established session keys. Assume adversary $\mathcal{A}$  has  eavesdropped the transferred messages $T_{BA},T_{CA},T_{AB},T_{CB},T_{AC}$ and $T_{BC}$.
From the values $T_{BA},T_{AB},T_{CA},S_A$ and $S_B$, adversary $\mathcal{A}$ can compute $k=\hat{e}(T_{AB}, S_B)\hat{e}(T_{BA}, S_A)\hat{e}(T_{CA}, S_A) =\hat{e}(P,P)^{a+b+c}$ from which he can construct the session key. Thus the XCQ-11 protocol cannot provide forward secrecy.
\subsection{KCI attack by a common adversary}
Suppose that $A$'s long-term private key $S_A$ and $B$'s long-term private key $S_B$  have been compromised. Obviously, $\mathcal{A}$ is now able to impersonate the corrupted party to any other party. However, it is also desirable that knowledge of the full private key does not enable $\mathcal{A}$ to impersonate other entities to the corrupted party. Accordingly, in a three-party key agreement protocol, a KCI attack can be an attack whereby  $\mathcal{A}$, with $A$'s long-term private key  and $B$'s long-term private key at hand, attempts to establish a valid session key with $A$ and $B$ by masquerading as another legitimate entity (say $C$).

A detailed description of KCI attack by a common adversary against the XCQ-11 protocol is outlined below ($\mathcal{A}(C)$ denotes that $\mathcal{A}$ is impersonating $C$).
\begin{enumerate}

\item[(1)] $A, B, \mathcal{A}(C)$: choose $a, b, c' \in Z_q^*$.
\item[(2)] $A \rightarrow B: T_{AB}= a(upk_B + H_2(upk_B)Q_B),$

\noindent  $ A \rightarrow \mathcal{A}(C): T_{AC}= a(upk_C + H_2(upk_C)Q_C),$

\noindent  $ B \rightarrow A: T_{BA}= b(upk_A + H_2(upk_A)Q_A),$

\noindent  $ B \rightarrow \mathcal{A}(C): T_{BC}= b(upk_C + H_2(upk_C)Q_C),$

\noindent  $ \mathcal{A}(C) \rightarrow A: T_{CA}= c'(upk_A + H_2(upk_A)Q_A),$

\noindent  $ \mathcal{A}(C) \rightarrow B: T_{CB}= c'(upk_B + H_2(upk_B)Q_B).$

\item[(3)]$A$ and $B$ compute the session key according to the protocol specification. $\mathcal{A}(C)$ computes the session key as follows.

$k_{\mathcal{A}(C)} =\hat{e}(P, P)^{c'} \hat{e}(T_{AB}, S_B)\hat{e}(T_{BA}, S_A)=  \hat{e}(P, P)^{a+b+c'}.$

$K=H_3(ID_A||ID_B||ID_C||upk_A||upk_B||upk_C||T_{AB}||T_{AC}||T_{BA}||T_{BC}||T_{CA}||T_{CB}||$

 $\qquad\hat{e}(P, P)^{a+b+c'})$
\end{enumerate}
So $\mathcal{A}$  successfully agrees a session key $K$ with entity $A$ and $B$ while $A$ and $B$ believes he is
sharing the key with entity $C$. Thus KCI attack by a common adversary is successful.
\subsection{An attack by an adversary who knows all users' secret values}
  An  adversary who knows all users' secret values can compute the session key of the XCQ-11 protocol with the following method.

  From the values $T_{AB},T_{AC},T_{BA},T_{BC},T_{CA},T_{CB},upk_A,upk_B, upk_C,q_A, q_B, q_C,x_A, x_B$ and $x_C$, the adversary can  compute
the following three points

  $aP= \frac{1}{q_B-q_C}(\frac{1}{x_B+H_2(upk_B)}T_{AB}-\frac{1}{x_C+H_2(upk_C)}T_{AC})$

      $\quad\ = \frac{1}{q_B-q_C}(\frac{1}{x_B+H_2(upk_B)}a(x_B + H_2(upk_B))Q_B$

       $\qquad\qquad-\frac{1}{x_C+H_2(upk_C)}a(x_C + H_2(upk_C))Q_C)$

       $\quad\ = \frac{1}{q_B-q_C}(aQ_B-aQ_C)$

       $\quad\ = \frac{1}{q_B-q_C}(q_B-q_C)aP$

 $bP= \frac{1}{q_A-q_C}(\frac{1}{x_A+H_2(upk_A)}T_{BA}-\frac{1}{x_C+H_2(upk_C)}T_{BC})$

 $cP= \frac{1}{q_A-q_B}(\frac{1}{x_A+H_2(upk_A)}T_{CA}-\frac{1}{x_B+H_2(upk_B)}T_{CB})$.

  Then the adversary can compute  $k=\hat{e}(aP+bP+cP, P)=\hat{e}(P,P)^{a+b+c}$ from which he can obtain the session key.

\subsection{Our improvement}
The reason why the XCQ-11 protocol can suffer from the above three attacks is that it lacks message origin authentication in the XCQ-11 protocol.  To make up security weaknesses, we give a simple improvement which uses signatures to achieve message origin authentication and has the same design idea as protocols \cite{Shim,HXG}.
 \leftmargini=3mm
\leftmarginii=5.5mm
\begin{itemize}
\setlength{\parskip}{-1pt}
\item  {\bf Setup}: This phase is the same as that in Section 3.1 except that a secure signature scheme from pairings  is chosen and $H_3$ is modified to $H_3:{\{0,1\}^*}^3 \times  \mathbb{G}_1^{6}\times \mathbb{G}_2  \rightarrow \{0,1\}^k$.

\item  {\bf  PartialKeyGen and UserKeyGen}: These two phases are the same as those in Section 3.1.

\item {\bf Key Agreement}: This phase is the same as that in Section 3.1 except that the following message flows and computations of a protocol run.
\begin{enumerate}
\setlength{\parskip}{0pt}
\item[(1)] $A, B, C$: choose $a, b, c \in Z_q^*$.
\item[(2)]
$A \rightarrow B,C: \{T_A= aP,\sigma_A$\}, where $\sigma_A$ is the signature on $T_A$ and $upk_A$ under $A$'s full private key $S_A$.

\noindent  $B \rightarrow A,C: \{T_B= bP,\sigma_B$\},  where $\sigma_B$ is the signature on $T_B$ and $upk_B$ under $B$'s full private key $S_B$.

\noindent  $C \rightarrow A,B: \{T_C= cP,\sigma_C$\}, where $\sigma_C$ is the signature on $T_C$ and $upk_C$ under $C$'s full private key $S_C$.
\item[(3)]
$A$ verifies the  validity of $\sigma_B$ and $\sigma_C$. If both are valid, $A$ computes $k_A=\hat{e}(T_B, T_C)^a$.

\noindent $B$ verifies the  validity of $\sigma_A$ and $\sigma_C$. If both are valid, $B$ computes $k_B=\hat{e}(T_A, T_C)^b$.

\noindent $C$ verifies the  validity of $\sigma_A$ and $\sigma_B$. If both are valid, $C$ computes $k_C=\hat{e}(T_A, T_B)^c$.
\end{enumerate}

\end{itemize}
After the protocol has finished, all three entities share the session key, which is computed as

$K=H_3(ID_A||ID_B||ID_C||upk_A||upk_B||upk_C||T_A||T_B||T_C||\hat{e}(P, P)^{abc})$.

With this modification, the improved protocol can withstand the above attacks. Reasons are easily described as follows.

The resulting session key of our improved protocol is independent of the participants' full private keys as the full private keys are used only to generate signatures. That is to say,  compromising the full private keys of all participants is no help to compute the session key. Hence, the improved protocol provides perfect forward secrecy and resist the attack described in Section 3.4.  Furthermore, an adversary who wants to impersonate a user must generate a correct signature, however, he cannot generate a correct signature without the user's full private key. Thus the improved protocol can resist KCI attack by a common adversary.
\section{The XCL-12 protocol and its analysis and improvement}
In this section, we first review the XCL-12 protocol  \cite{XCL}, then  show that the XCL-12 protocol is  vulnerable to two types of  KCI attacks, and finally propose an efficient countermeasure to resist these attacks.
\subsection{Review of  the  XCL-12 protocol}
The XCL-12 protocol \cite{XCL} is described as follows.
\leftmargini=3mm
\leftmarginii=5.5mm
\begin{itemize}
\setlength{\parskip}{-1pt}
\item  {\bf Setup}: Given a security parameter $k\in \mathbb{Z}$, the algorithm works as follows.
\begin{enumerate}
\setlength{\parskip}{-1pt}
\item[(1)] It runs the parameter generator on input $k$ to generate a prime $q$, two groups $\mathbb{G}_1, \mathbb{G}_2$ of prime order $q$, a generator $P$ of $\mathbb{G}_1$, and an admissible pairing $\hat{e}: \mathbb{G}_1 \times\mathbb{G}_1 \rightarrow \mathbb{G}_2$.

\item[(2)]  It chooses a master-key $x\in Z_q^*$ and computes $P_0= xP$.

\item[(3)]  It chooses two cryptographic secure hash functions $H_1: \{0,1\}^* \times  \mathbb{G}_1 \rightarrow \mathbb{Z}_q^*$  and $H_2:{\{0,1\}^*}^3 \times  \mathbb{G}_1^{10}\times \mathbb{G}_2^2  \rightarrow \{0,1\}^k$. Finally the KGC's
master-key $x$ is kept secret and the system parameters $\{q,\mathbb{G}_1,\mathbb{G}_2, \hat{e}, P, P_0, H_1, H_2\}$ are published.
\end{enumerate}

\item {\bf PartialKeyGen}: Given a user's identity $ID_U \in \{0,1\}^*$, KGC first chooses at random $r_U\in Z_q^*$, and computes $R_U = r_UP, h = H_1(ID_U||R_U)$, and $s_U = (r_U + hx)^{-1}$. It then sets this user's partial private key $\{s_U, R_U\}$ and transmits it to user $ID_U$ secretly.

It is easy to see that user $ID_U$ can validate his/her partial private key by checking whether the equation $s_U (R_U +H_1(ID_U||
R_U)P_0) = P$ holds. The partial key is valid if the equation holds, and vice versa.

\item {\bf UserKeyGen}: User  $ID_U$  picks randomly $x_U\in Z_q^*$ as his/her user secret key $usk_U$, and computes  his/her public key as $upk_U = x_UP$.

\item {\bf Key Agreement}: Assume that an entity $A$ with identity $ID_A$ has full private key $(s_A,R_A,x_A)$ and public key $upk_A$, an entity $B$ with identity $ID_B$ has private key $(s_B,R_B,x_B)$ and public key $upk_B$, and an entity $C$ with identity $ID_C$ has private key $(s_C,R_C,x_C)$ and public key $upk_C$. The message flows and computations of a protocol run are described below.
\begin{enumerate}
\setlength{\parskip}{1pt}
\item[(1)] $A, B, C$: choose $a, b, c \in Z_q^*$.
\item[(2)]
$A \rightarrow B, C: \{ID_A, upk_A, R_A\}$

\noindent  $B \rightarrow A: \{ID_B,upk_B,R_B,T_{BA}= b(R_A + H_1(ID_A ||R_A)P_0)\}$

\noindent  $C \rightarrow A: \{ID_C,upk_C,R_C,T_{CA}= c(R_A + H_1(ID_A ||R_A)P_0)\}$

\noindent  $A \rightarrow B: T_{AB} = a(R_B + H_1(ID_B|| R_B)P_0)$

\noindent  $A \rightarrow C: T_{AC} = a(R_C + H_1(ID_C||R_C)P_0)$

\noindent  $B \rightarrow C: \{ID_B, upk_B, R_B\}$

\noindent  $C \rightarrow B: \{ID_C , upk_C , R_C , T_{CB} = c(R_B + H_1(ID_B ||R_B)P_0)\}$

\noindent  $B \rightarrow C: T_{BC} = b(R_C + H_1(ID_C|| R_C )P_0).$

\item[(3)]$A$ computes:

$ k^1_{ABC} = aP + s_AT_{BA} + s_AT_{CA} = aP + bP + cP = (a + b + c)P$

$ k^2_{ABC} = \hat{e}(s_AT_{BA}, s_AT_{CA})^a = \hat{e}(bP, cP)^a = \hat{e}(P, P)^{abc}$

$k^3_{ABC} = \hat{e}(upk_B, upk_C )^{x_A} = \hat{e}(P, P)^{x_Ax_Bx_C}.$

\noindent $B$ computes:

 $k^1_{ABC} = bP + s_BT_{AB} + s_BT_{CB} = bP + aP + cP = (a + b + c)P$

 $k^2_{ABC} = \hat{e}(s_BT_{AB}, s_BT_{CB})^b = \hat{e}(aP, cP)^b = \hat{e}(P, P)^{abc}$

$k^3_{ABC} = \hat{e}(upk_A, upk_C )^{x_B} = \hat{e}(P, P)^{x_Ax_Bx_C}.$

\noindent$C$ computes:

 $k^1_{ABC} =cP + s_CT_{AC} + s_CT_{BC} = cP + aP + bP = (a + b + c)P$

 $k^2_{ABC} = \hat{e}(s_CT_{AC}, s_CT_{BC})^c = \hat{e}(aP, bP)^c = \hat{e}(P, P)^{abc}$

 $k^3_{ABC} = \hat{e}(upk_A, upk_B )^{x_C} = \hat{e}(P, P)^{x_Ax_Bx_C}.$

\end{enumerate}
\end{itemize}

After the protocol has finished, all three entities share the session key, which is computed as

$K=H_2(ID_A||ID_B||ID_C||upk_A||upk_B||upk_C||T_{AB}||T_{AC}||T_{BA}||T_{BC}||T_{CA}||T_{CB}||$

$ \qquad(a + b + c)P||\hat{e}(P, P)^{abc}||\hat{e}(P, P)^{x_Ax_Bx_C})$.
\subsection{ KCI attack by a malicious KGC}
Suppose the full private key $(s_A,R_A,x_A)$ of an entity $A$ is compromised by a malicious KGC (say $\mathcal{E}$). Obviously, $\mathcal{E}$ is now able to impersonate the corrupted party to any other party. However, it is also desirable that knowledge of the full private key does not enable $\mathcal{E}$ to impersonate other entities to the corrupted party.  Accordingly, in a three-party key agreement protocol, a KCI attack can be an attack whereby $\mathcal{E}$, with $A$'s long-term private key at hand, attempts to establish a valid session key with $A$ and $B$ by masquerading as another legitimate entity (say $C$).

A detailed description of KCI attack by a malicious attack against the XCL-12 protocol  is outlined below ($\mathcal{E}(C)$ denotes that $\mathcal{E}$ is impersonating $C$). We note that a malicious KGC knows the partial key $(s_C,R_C)$ of $C$ since the user's partial key is generated by him, however, he cannot know the secret value $x_C$ of $C$.
\begin{enumerate}
\setlength{\parskip}{1pt}
\item[(1)] $A,B , \mathcal{E}(C)$: choose $a, b, c' \in Z_q^*$.

\item[(2)]
 $A \rightarrow B, \mathcal{E}(C): \{ID_A, upk_A, R_A\}$

\noindent $B \rightarrow A: \{ID_B, upk_B, R_B, T_{BA}= b(R_A + H_1(ID_A ||R_A)P_0) \}$

\noindent $\mathcal{E}(C) \rightarrow A: \{ID_C, upk_C, R_C, T_{CA}= c'(R_A + H_1(ID_A ||R_A)P_0)\}$

\noindent $A \rightarrow B: T_{AB} = a(R_B + H_1(ID_B|| R_B)P_0)$

\noindent $A \rightarrow \mathcal{E}(C): T_{AC} = a(R_C + H_1(ID_C||R_C)P_0)$

\noindent  $B \rightarrow \mathcal{E}(C): \{ID_B, upk_B, R_B\}$

\noindent $\mathcal{E}(C) \rightarrow B: \{ID_C , upk_C , R_C , T_{CB} = c'(R_B + H_1(ID_B ||R_B)P_0)\}$

\noindent $B \rightarrow \mathcal{E}(C): T_{BC} = b(R_C + H_1(ID_C|| R_C )P_0)$.

\item[(3)] $A$ and $B$ compute the session key according to the protocol specification. $\mathcal{E}(C)$ computes the session key as follows.

 $k^1_{ABC} = c'P + s_CT_{AC} + s_AT_{BA} = c'P + aP + bP = (a + b + c')P$

 $k^2_{ABC} = \hat{e}(s_CT_{AC}, s_AT_{BA})^{c'}= \hat{e}(aP, bP)^{c'} = \hat{e}(P, P)^{abc'}$

 $k^3_{ABC} = \hat{e}(upk_B, upk_C )^{x_A} = \hat{e}(P, P)^{x_Ax_Bx_C}$

  $K=H_2(ID_A||ID_B||ID_C||upk_A||upk_B||upk_C||T_{AB}||T_{AC}||T_{BA}||T_{BC}||T_{CA} ||T_{CB}||$

  $\qquad(a + b + c')P||\hat{e}(P, P)^{abc'}||\hat{e}(P, P)^{x_Ax_Bx_C})$.
\end{enumerate}
 So $\mathcal{E}$  successfully agrees a session key $K$ with entity $A$ and $B$ while $A$ and $B$ believes he is
sharing the key with entity $C$. Thus KCI attack by a malicious KGC is successful.

\subsection{ KCI Attack by a common adversary}

Suppose that $A$'s full private key $(s_A,R_A,x_A)$  and $B$'s full private key $(s_B,R_B,x_B)$   have been compromised. Obviously, $\mathcal{A}$ is now able to impersonate the corrupted party to any other party. However, it is also desirable that knowledge of the full private key does not enable $\mathcal{A}$ to impersonate other entities to the corrupted party. Accordingly, in a three-party key agreement protocol, a KCI attack can be an attack whereby  $\mathcal{A}$, with $A$'s long-term private key  and $B$'s long-term private key at hand, attempts to establish a valid session key with $A$ and $B$ by masquerading as another legitimate entity (say $C$).

A detailed description of KCI attack by a common adversary against the XCL-12 protocol is outlined below ($\mathcal{A}(C)$ denotes that $\mathcal{A}$ is impersonating $C$).
\begin{enumerate}
\setlength{\parskip}{1pt}
\item[(1)] $A,B , \mathcal{A}(C)$: choose $a, b, c'' \in Z_q^*$.

\item[(2)]
 $A \rightarrow B, \mathcal{A}(C): \{ID_A, upk_A, R_A\}$

\noindent $B \rightarrow A: \{ID_B, upk_B, R_B, T_{BA}= b(R_A + H_1(ID_A ||R_A)P_0) \}$

\noindent $\mathcal{A}(C) \rightarrow A: \{ID_C, upk_C, R_C, T_{CA}= c''(R_A + H_1(ID_A ||R_A)P_0)\}$

\noindent $A \rightarrow B: T_{AB} = a(R_B + H_1(ID_B|| R_B)P_0)$

\noindent $A \rightarrow \mathcal{A}(C): T_{AC} = a(R_C + H_1(ID_C||R_C)P_0)$

\noindent  $B \rightarrow \mathcal{A}(C): \{ID_B, upk_B, R_B\}$

\noindent $\mathcal{A}(C) \rightarrow B: \{ID_C , upk_C , R_C , T_{CB} = c''(R_B + H_1(ID_B ||R_B)P_0)\}$

\noindent $B \rightarrow \mathcal{A}(C): T_{BC} = b(R_C + H_1(ID_C|| R_C )P_0)$.

\item[(3)] $A$ and $B$ compute the session key according to the protocol specification. $\mathcal{A}(C)$ computes the session key as follows.

 $k^1_{ABC} = c''P + s_BT_{AB} + s_AT_{BA} = c''P + aP + bP = (a + b + c'')P$

 $k^2_{ABC} = \hat{e}(s_BT_{AB}, s_AT_{BA})^{c''}= \hat{e}(aP, bP)^{c''} = \hat{e}(P, P)^{abc''}$

 $k^3_{ABC} = \hat{e}(upk_B, upk_C )^{x_A} = \hat{e}(P, P)^{x_Ax_Bx_C}$

  $K=H_2(ID_A||ID_B||ID_C||upk_A||upk_B||upk_C||T_{AB}||T_{AC}||T_{BA}||T_{BC}||T_{CA} ||T_{CB}||$

  $\qquad(a + b + c')P||\hat{e}(P, P)^{abc'}||\hat{e}(P, P)^{x_Ax_Bx_C})$.
\end{enumerate}
 So $\mathcal{A}$  successfully agrees a session key $K$ with entity $A$ and $B$ while $A$ and $B$ believes he is
sharing the key with entity $C$. Thus KCI attack  by a common adversary is successful.

\subsection{Our improvement}
 Informally saying, the XCQ-12 protocol cannot resist two types of  KCI attacks is because the inappropriate design of shared values $k^1_{ABC}, k^2_{ABC}$ and $k^3_{ABC}$ makes that the session key does not depend on all three  parties' partial private keys, secret values, and ephemeral secrets. To make up the security weaknesses, we give an efficient improvement as follows which modifies the three shared values.
 \leftmargini=3mm
\leftmarginii=5.5mm
\begin{itemize}
\setlength{\parskip}{1pt}
\item  {\bf Setup,PartialKeyGen and UserKeyGen}: These three phases are the same as those in Section 4.1.

\item {\bf Key Agreement}: This phase is the same as that in Section 4.1 except that the following computations.

$A$ computes

$ k^1_{ABC} = aP + s_AT_{BA} + s_AT_{CA} = aP + bP + cP = (a + b + c)P$

$k^2_{ABC}= \hat{e}(s_AT_{BA}+R_B + H_1(ID_B|| R_B)P_0, s_AT_{CA}+R_C +H_1(ID_C|| R_C)P_0)^{a +s_A^{-1}} $

 $\qquad \ \ = \hat{e}(P, P)^{(a+s_A^{-1})(b+s_B^{-1})(c+s_C^{-1})}$

$k^3_{ABC} = \hat{e}(s_AT_{BA}+upk_B, s_AT_{CA}+upk_C )^{a +x_A}= \hat{e}(P, P)^{(a+x_A)(b+x_B)(c+x_C)}$

\noindent $B$ computes

 $k^1_{ABC} = bP + s_BT_{AB} + s_BT_{CB} = bP + aP + cP = (a + b + c)P$

 $ k^2_{ABC} = \hat{e}(s_BT_{AB}+R_A + H_1(ID_A ||R_A)P_0, s_BT_{CB}+R_C +H_1(ID_C|| R_C)P_0)^{b+s_B^{-1}}$

 $\qquad\ \   = \hat{e}(P, P)^{(a+s_A^{-1})(b+s_B^{-1})(c+s_C^{-1})}$

 $ k^3_{ABC} = \hat{e}(s_BT_{AB}+upk_A, s_BT_{CB}+upk_C )^{b+x_B}= \hat{e}(P, P)^{(a+x_A)(b+x_B)(c+x_C)}$

\noindent$C$ computes

 $k^1_{ABC} =cP + s_CT_{AC} + s_CT_{BC} = cP + aP + bP = (a + b + c)P$

 $k^2_{ABC} = \hat{e}(s_CT_{AC}+R_A + H_1(ID_A ||R_A)P_0, s_CT_{BC}+R_B +H_1(ID_B|| R_B)P_0)^{c+s_C^{-1}} $

 $\qquad\ \   =\hat{e}(P, P)^{(a+s_A^{-1})(b+s_B^{-1})(c+s_C^{-1})}$

 $k^3_{ABC} = \hat{e}(s_CT_{AC}+upk_A, s_CT_{BC}+upk_B )^{c+x_C}= \hat{e}(P, P)^{(a+x_A)(b+x_B)(c+x_C)}.$
\end{itemize}
After the protocol has finished, all three entities share the session key, which is computed as

$K=H_2(ID_A||ID_B||ID_C||upk_A||upk_B||upk_C||T_{AB}||T_{AC}||T_{BA} ||T_{BC}|| T_{CA}||T_{CB}||$

 $\qquad \ \   (a + b + c)P||\hat{e}(P, P)^{(a+s_A^{-1})(b+s_B^{-1})(c+s_C^{-1})}||\hat{e}(P, P)^{(a+x_A)(b+x_B)(c+x_C)})$.

With this modification, the improved protocol can withstand  two types of KCI attacks   due to the following reasons.

As we know, a malicious KGC can know partial private keys $(s_A,R_A)$,$(s_B,R_B)$ and $(s_C,R_C)$. Suppose the full private key $(s_A,R_A,x_A)$ of an entity $A$ is compromised by a malicious KGC. Then if he want to  impersonate  $C$ to $A$ and $B$, he would have to compute $k^3_{ABC}=\hat{e}(s_CT_{AC}+x_AP, s_CT_{BC}+upk_B )^{c'+x_C}$. However, without the knowledge of $a$ and $x_C$, the malicious KGC cannot compute $k^3_{ABC}$ since he must know $b$ and $x_B$ which is not permitted.

Suppose that $A$'s full private key $(s_A,R_A,x_A)$  and $B$'s full private key $(s_B,R_B,x_B)$ have been compromised by an adversary $\mathcal{A}$. Then if he want to  impersonate  $C$ to $A$ and $B$, he would have to compute $k^2_{ABC} = \hat{e}(s_BT_{AB}+s_A^{-1}P, s_AT_{BA}+s_A^{-1}P)^{c''+s_C^{-1}}$ and $k^3_{ABC}=\hat{e}(s_BT_{AB}+x_AP, s_AT_{BA}+x_BP )^{c''+x_C}$. However, without the knowledge of $a$ and $b$, $\mathcal{A}$ cannot compute $k^2_{ABC}$  and $k^3_{ABC}$ since he must know $s_C$ and $x_C$ which is not permitted.

Furthermore, our improved protocol is as efficient as the XCQ-12 protocol since only 4 point additions are increased.

\section{Conclusion}
In this paper, we have indicated that Xiong et al.'s protocol \cite{XCQ} suffers from FS attack, KCI attack and an attack by adversaries who know all users' secret values, and proposed a simple improvement to remove these flaws.  We also have indicated that Xiong et al.'s  protocol \cite{XCL} cannot resist two types of  KCI  attacks and  proposed an efficient improvement to remove these flaws.

\section*{Acknowledgement}
This work is supported by NSFC (Grant Nos. 61272057, 61202434, 61170270, 61100203, 61003286, 61121061), the Fundamental Research Funds  for the Central Universities (Grant Nos.  2012RC0612, 2011YB01).

\bibliographystyle{unsrt}

\end{document}